# Real time display with the ferrolens of homogeneous magnetic fields

Emmanouil Markoulakis[a*], Timm Vanderelli[b], Lambros Frantzeskakis[a]

[a] *Department of Electronic Engineering, Computer Technology Informatics & Electronic Devices Laboratory, Hellenic Mediterranean University, Romanou 3, Chania 73133, Crete, Greece.*
[b] *Ferrocell USA, Research and Development division, 739 Route 259, Ligonier, PA 15658, USA.*
[*] *Corresponding author: E-mail address:* <u>markoul@hmu.gr</u> *(E. Markoulakis)*

## Abstract

In our previous published research we have studied the applications of the ferrolens for the observation and qualitative analysis of non-homogeneous magnetic fields. Latest developments over the last few years of the ferrolens increased multifold the sensitivity of this device allowing it to display magnetic fields as low as 10mT in strength and therefore it is possible now to observe also the homogeneous (i.e. straight parallel magnetic flux lines of uniform density in space) magnetic field existing inside air solenoids and between N-S poles of two attracting separated permanent magnets. We present and analyze herein these novel observations of homogeneous magnetic fields with the ferrolens and as a potential new application of this device. The ferrolens can now display the projected magnetic field on air from a distance without needing to be in physical contact with the field source. Experiments were carried out to demonstrate these new capabilities of the ferrolens as a nanomagnetic flux viewer, real-time physical device and scientific qualitative tool. We compare the geometry of magnetic fields of condensed matter ferromagnets with that of magnetic fields inside electrical air solenoids. Specifically, the homogeneous field between two N-S attracting magnets at a distance as well as inside air solenoids and Helmholtz coils and also the non-homogeneous total field of single dipole permanent magnets were observed using a latest generation very sensitive ferrolens. The unique feature of this magneto optic fluid thin film physical device is that it can display discrete magnetic flux lines of the macroscopic field. Also we show how the ferrolens can be used to detect qualitatively the symmetry center of a non ideal homogeneous magnetic field.

*Keywords:* physics instrumentation, quantum magnetic flux viewer, magnetic field geometry and topology, ferromagnetism, homogeneous magnetic fields, Helmholtz coils.

## 1. Introduction

For the last fourteen years numerous researches have studied and published research about the ferrolens and magnetic fields as well magneto-optics driven light polarization related research using the ferrolens [1–8] [10–12]. We herein concentrate exclusively on the research of magnetic fields using the ferrolens as a real time magnetic field viewer physical device. Our primary goal herein is to present and analyze the results from our new novel observations and experiments of homogeneous magnetic fields with the ferrolens as a potential new proposed application for this device. The operational and characteristics description of the device and field display mechanism as well as the quantum and microscopic physics aspects of this device were extensively described, analyzed and based also on independent sources cited research and literature done over the last decades on magnetic thin films and ferrofluid in general and can all be found inside our previous published related research [1][2][3][4]. The results of the observation of non-homogeneous magnetic fields were presented in our previous work like this of a high-voltage radio antenna [1], permanent dipole magnets and magnetic arrays and lattices [2][3] and macroscopic synthetic magnetic monopole array [4]. Nevertheless, we used a few of our previous material, absolutely necessary for the comparison of the display of the ferrolens between homogeneous and non-homogeneous magnetic fields.

## 2. Materials and methods

A ferrolens[1,2] [1–6] [7–12] consists of two optical quality glass disks separated by a thin-film of superparamagnetic fluid. This fluid is comprised of nano-sized magnetite $Fe_3O_4$ particles less than 10nm in size average, coated with a polar surfactant oleic acid and suspended in a light hydrocarbon-based colloidal solution, commonly known as Ferrofluid *(EFH1[3] product was used)* which is however further heavily diluted and refined using a centrifuge machine in the case of the ferrolens's thin-film which is about 10μm thick *(latest generation)* and by using additional hydrocarbon based solvent *(e.g. benzene)*. When a ferrolens is irradiated with visible artificial light and under the influence of magnetism, it will display colored paths that are shaped by the applied magnetic field and aligned with it. The colored paths are due to the light scattering effect on the formed nanochains aligned to the external magnetic field. The whole imprint of the incident magnetic field on the ferrolens is compressed inside the thickness of the thin film. The ferrolens therefore is displaying a 2D compressed cross-section imprint in real time of the actual 3D Euclidean space incident to the lens magnetic field and different observation angles for the field can be adjusted. Because the finite thickness of the thin film used inside the lens, it allows the display of the field by the ferrolens to be not completely flat 2D but some depth of field information is retained resulting to a tunnel-effect display of the field [2][3]. Also the display of the field by the ferrolens features magnetic transparency allowing us to see the field across pole to pole of the magnet without the bulk of the magnet obscuring our view. All these we will demonstrate with our findings next. In the experiments several shapes standard neodymium magnets (Nd) were used. The ferrolens we used was the latest generation mini-Ferrocell with 10mT sensitivity and 10μm thin film thickness. The increase of the magneto optic response sensitivity of the ferrolens is a result of the progressive reduction of the ferrolenses's thin film thickness achieved over the last years but also due to the increase of the ferrofluid dilution factor (i.e. less volumetric concentration percentage of magnetite nanoparticles inside the carbon-based colloidal solution). Thus from an initial about 70mT sensitivity of the thin film with 50μm film thickness we improved the sensitivity to 10mT and 10μm thin film thickness. The total thickness of the mini-Ferrocell (i.e. model name) ultra sensitive ferrolens device used in the experiments including its two optical grade glasses is 1mm. Of course this increase in sensitivity comes with a drawback, the old 25-50 μm models, 4-8mm total thickness, could display the field of Nd magnets in contact with the ferrolens surface, of residual magnetism $B_r$ of their bulk up to 1.3 T without destroying the device even for prolonged periods of exposure (The so called "burned out" effect, permanent damage to a 6mm total thickness ferrolens after a week of continuous exposure to a strong Nd magnet placed under its surface, is demonstrated using microscopy in this video https://tinyurl.com/3ap4xr5w , *please read description of the video*).

The ultra sensitive ferrolens device (mini-Ferrocell) we used herein is designed more for probing a magnetic field on air without physical contact with the magnet therefore ideal for probing magnetic fields inside air solenoids and Helmholtz coils and other similar magnetic sources. Maximum field strength on the thin film for the mini-Ferrocell for prolonged exposure periods should not exceed 80mT. Another limitation factor affecting the longevity of the ferrolens devices in general besides prolonged exposure in strong magnetic fields (i.e. suggested use, not more than 2-5 min per observation period depending the device thickness and sensitivity and incident magnetic field strength) that will progressively damage the device, is the thin film evaporation with time. A normal use and optimal stored when not in use in a cool and dry place like a shelf (do not refrigerate) ferrolens can last for years but eventually it will evaporate specially on a hot ambient environment despite the fact that the ferrolens is sealed on its perimeter with a transparent ultra-violet light cured adhesive.

The Helmholtz coils used in the experiments were specifically calculated and constructed by us for the requirements of this research. At a measured value $B_c$ of 22mT field strength at their center, is adequate to trigger the display of the ferrolenses used. If demanded, they can draw safely about 150W of electrical power and besides wood other techniques and materials were used for their frame like 3D printing which can withstand temperatures without deforming up to 110 C°. The AWG 18 and AWG 20, magnet wire used in the construction was rated at 200C°. The Helmholtz coils were air cooled with a fan during operation which is necessary since without any cooling mechanism applied, temperature can rise above ambient in a rate up to +12 C° per minute. Operational periods lasted no more than 2-3 min per experimental session in order to ensure a control experiment.

For more technical details on the materials and methods used please see the supplementary material and appendix of this research *(see reference [13])*.

## 3. Results and discussion

### 3.1 Field observation results of ferromagnets

The basic experimental setup used of the two Nd magnets illustrated in **fig.1** *(see Appendix for technical details [13])*.

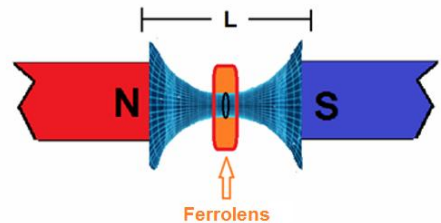

**Fig. 1** A ferrolens is placed at mid distance of the direct N-S field of two attracting permanent magnets kept at L distance apart. The lens surface is perpendicular to the N-S axis of the field therefore perpendicular to the magnetic lines of force. A 2D cross-section of the field is therefore displayed in real time on the ferrolens surface.

**Experiment A - For N-S axis of two attracting magnets perpendicular to the ferrolens glass surface (homogeneous field):**
We see clearly in fig.2a the magnetic flux lines are going straight without curl as in the case of the inside field of an air solenoid fig.2b shown by the ferrolens**,** through the formed funnel (like a wormhole) from one pole to the opposite pole as predicted by the well known theory for the direct homogeneous and straight magnetic flux lines fields formed between separated N-S poles of two magnets and also for the field inside an air solenoid.

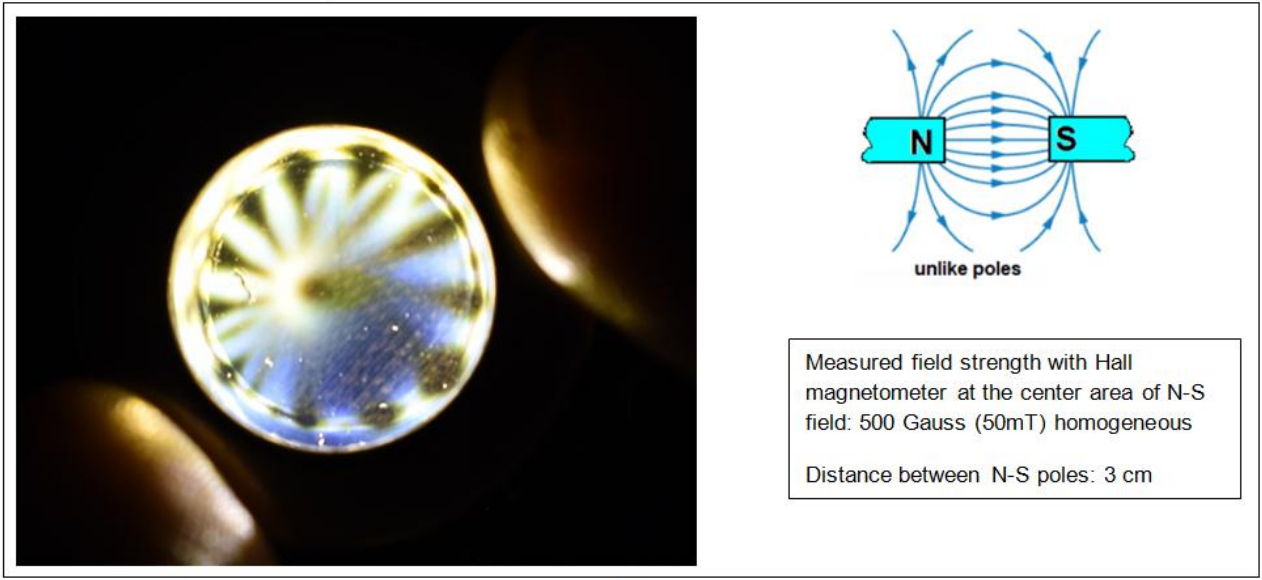
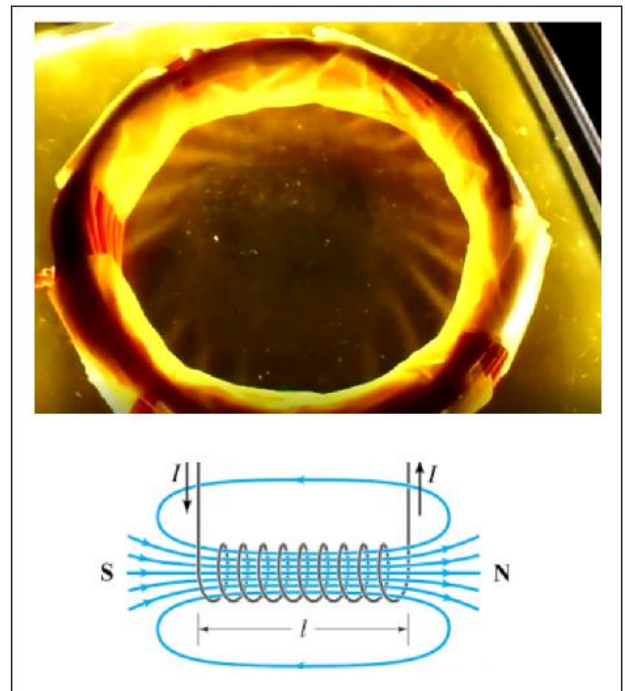
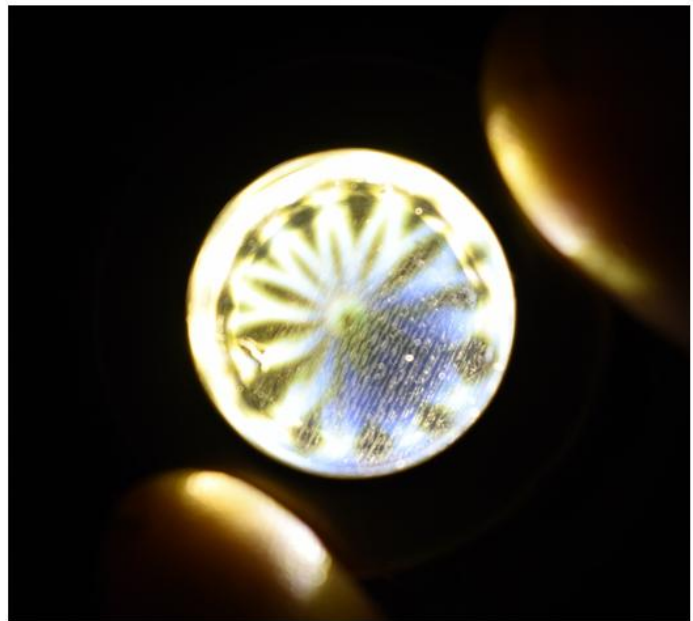
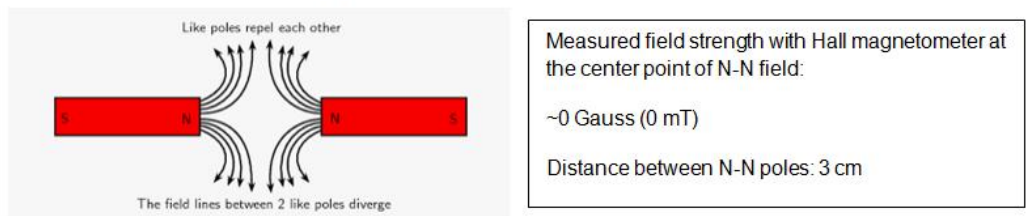

**Fig. 2 (a)** Direct N-S homogeneous field on air between N-S poles of two separated apart attracting magnets **(b)** Straight magnetic flux lines of the homogeneous field inside an air solenoid [4] **(c)** Direct N-N field on air between N-N poles of two separated apart magnets. The black region of previous photo fig.2a of the N-S case at the center becomes now white which means no magnetic flux is passing through the other pole N-N axis of magnets due the repulsion. The magnetic lines are deflected and the horn funnel is closed. Notice, that in the case of N-N repelling poles fig.2c, the flux lines shown on the ferrolens are bend and displayed less dark than at fig.2a of the N-S attracting poles. The lighter color is due the deflection of the lines.

**Experiment B - For N-N axis of two repulsing magnets perpendicular to the ferrolens glass surface (non-homogeneous field):**

Initially it seems that fig.2a of the attracting N-S poles field display we obtained with the ferrolens, is identical to this photo here of fig.2c of the N-N repulsing poles of two magnets. But there is a huge difference when we look more closely. In the case of fig. 2a of the attracting poles N-S, in the center of the tunnel path we see a relatively large black hole at the center that indicates that dense magnetic lines are passing through. In contrast in fig.2c of the repelling N-N poles this same region is now displayed mainly as a white luminous disk at the center of the ferrolens therefore indicating that there is zero magnetic flux passing through due the repulsion, towards the opposite pole magnet. The funnel path is closing up. As you approach even more close together the two N-N poles, this region will become completely luminous white whereas in **Experiment –A** of the display of the ferrolens of the attracting N-S poles, we see as you approach the two N-S poles, this same region will spread increasingly larger black in diameter which means that now more and more magnetic flux lines are passing through the funnel path formed between the two opposite attracting poles of the two magnets. Therefore in **Experiment–A** the funnel path is open in contrast to **Experiment –B** where the funnel is closed. Notice also, that in the case of N-N repelling poles, see fig.2c, the flux lines shown on the ferrolens are bend and displayed less dark than at fig.2a of the N-S attracting poles. The lighter color is due the deflection of the magnetic lines.

All above observations of **Experiment–B** confirm the theoretical predictions.

We can this see this also very clearly on the following movie, recorded from the experiments of the direct N-N repulsing field of two magnets each pole positioned on either side of a multi-color lid ferrolens, where we see as the magnets are approaching to each other thus the repulsion is getting stronger, the displayed funnel path (shown as a black hole on the lens) is closing due the deflection of the magnetic flux lines until it is totally closed (i.e. black hole disappears). Meaning, no magnetic lines of force are passing from one pole to the other and all lines are deflected due repulsion. We can also clearly see in this movie the deflected magnetic lines of force also depicted in the cartoon illustration of fig.2c, https://tinyurl.com/9569ys9w  (movie).

**When we compare** the above obtained results of Experiment-A and Experiment-B with these of the non-homogeneous total magnetic fields of **single dipole permanent magnets** [2][3] displayed by the ferrolens, see fig.3&4, the differennce is immidiatelly made visible and complies with what is predicted by theory. In fig.3 illustration, we present a possible compression effect of the incident magnetic field on the finite thickness thin-film magnetic optic sensor of the lens since the magnetic nanochains interacting inside the fluid solution of the thin film are subject of mechanical stress and pressure.

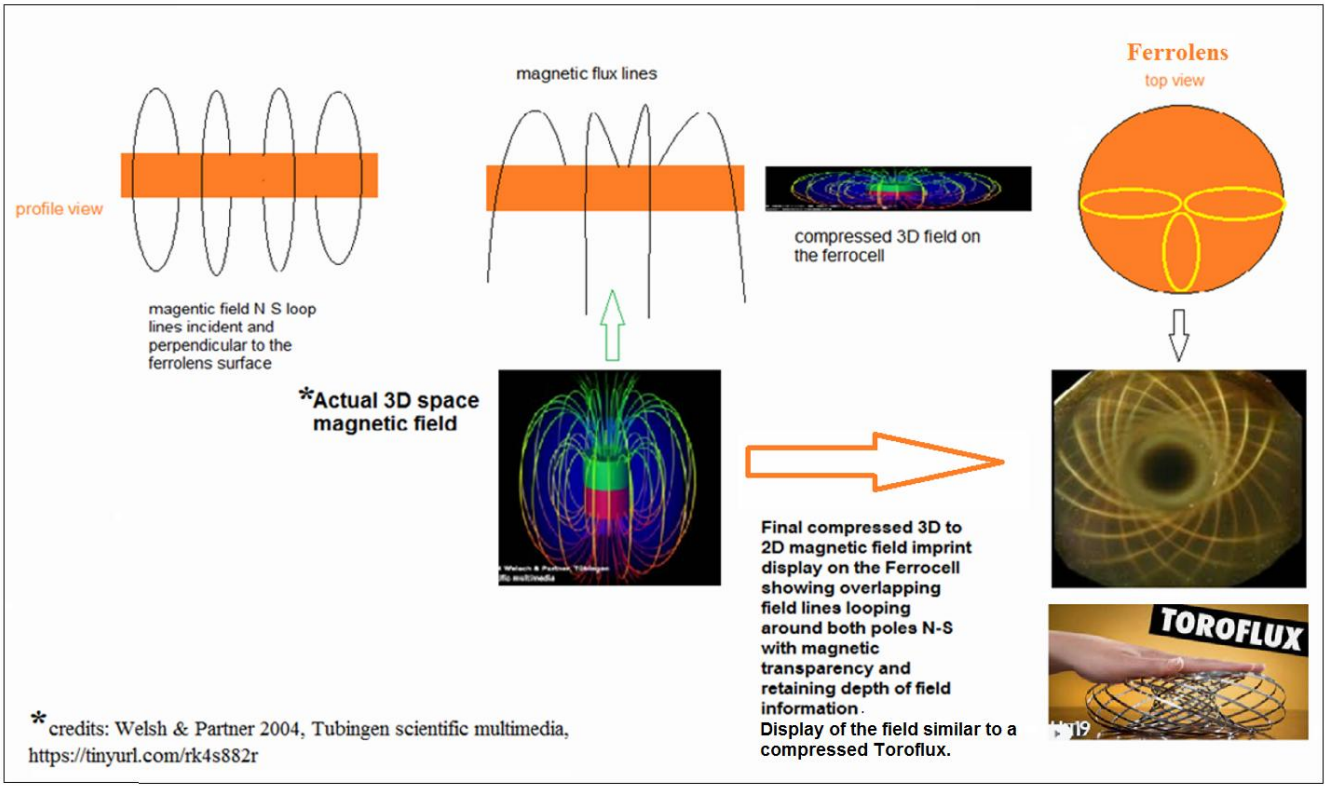

**Fig. 3** Illustration of the Ferrolens display of the actual 3D field and magnetic flux of ferromagnets as a possible compression effect of the incident magnetic field on the finite thickness thin-film magnetic optic sensor of the lens since the formed magnetic nanochains interacting inside the fluid solution of the thin film are subject of mechanical stress and pressure.

Also, electromagnets (i.e. solenoids with a ferromagnetic or ferrite core) from all of our experiments, geometrically generate the same field display on the ferrolens as single permanent ferromagnets as shown in fig.3 & fig.4.

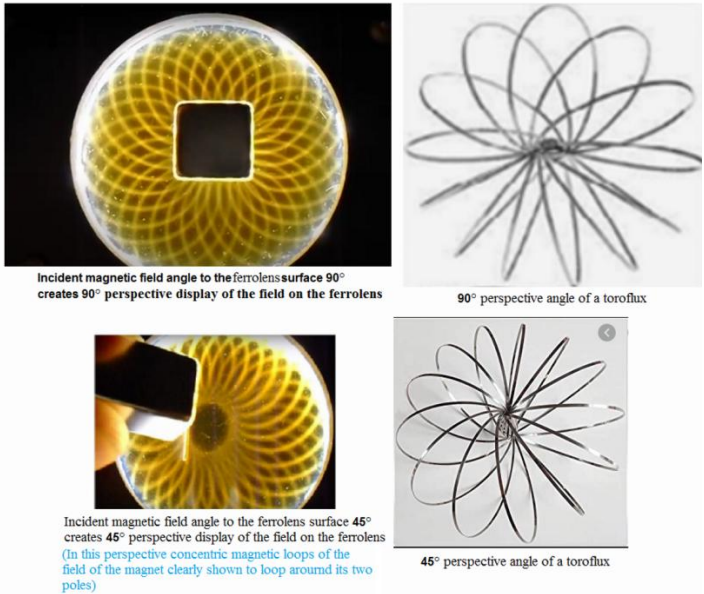

**Fig.4** Different incident magnetic field angles to the ferrolens surface allow different perspective display of the field.

In fig.4 different incident magnetic field angles to the surface of the ferrolens are compared and it is demonstrated how they correspond to the perspective display of the compressed field on the ferrolens. A 90° incident angle (i.e. pole axis of the magnet placed vertical to the lens surface, see fig.4 top) will produce a 90° perspective top view if the field whereas a 45° incident field angle will allow us to see the field of the magnet on the ferrolens from that perspective (see fig.4 bottom). Notice here, on the 45° perspective display of the field on fig.4, how clearly it is made visible, the magnetic N-S flux lines are looping in concentric circles around the two poles of the magnet displayed by the ferrolens with magnetic transparency and depth of field information (https://tinyurl.com/uke22vje movie).

*3.2 Observation results of homogeneous field of Helmholtz coil*

**Experiment C –** For the ferrolens positioned at the center of a 37.5mm radius Helmholtz coil and its N-S axis perpendicular to the ferrolens glass surface (**homogeneous field**):

A Helmholtz coil illustrated in fig.5a, for the requirements of the experiment was designed and constructed using a 3D printed frame *(see supplementary material, Appendix, for technical details and field linearity measurements [13])*. As before, the same ultra sensitive ferrolens was used. The field at the center o the Helmholtz coil $B_c$ was calculated and

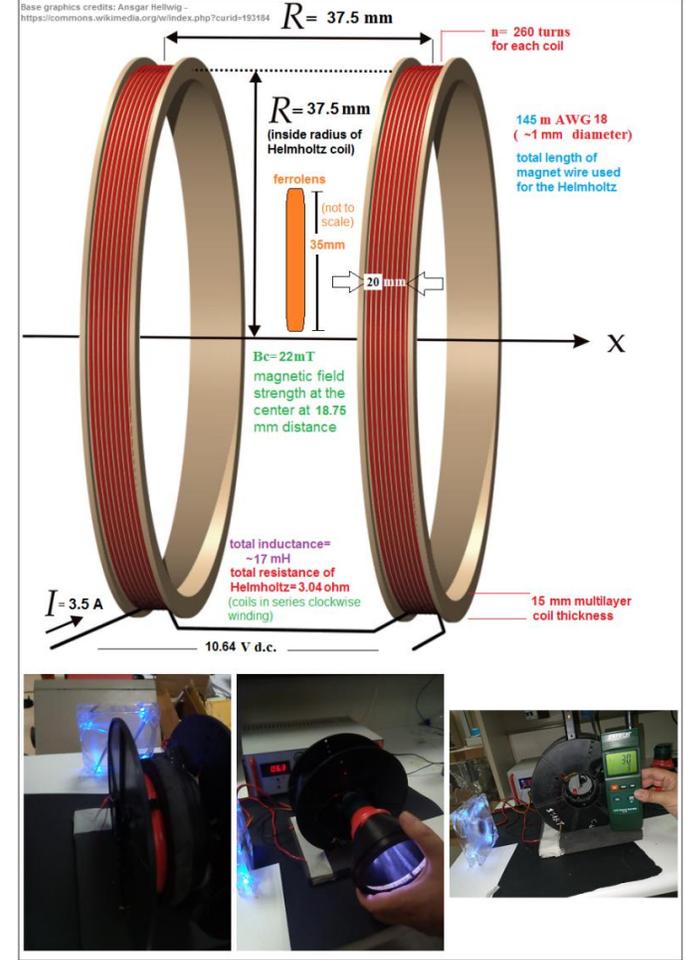

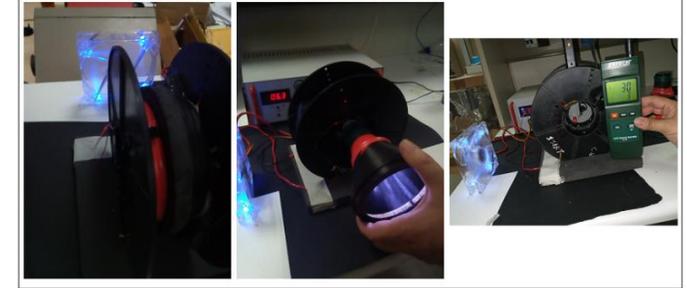

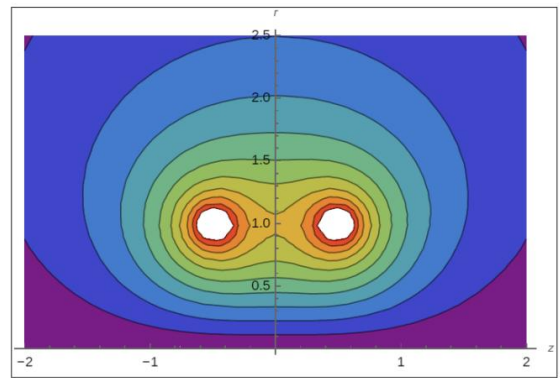

**Fig. 5 (a)** Helmholtz coil calculated for the experiment requirements. **(b)** Wolfram simulation of magnetic field curl of Helmholtz coils. Only about a 25% of the total volume of the field is ideal homogeneous (https://tinyurl.com/4awa5p37).

measured with a Hall magnetometer at 22mT field strength, double the sensitivity threshold of the ferrolens at 10mT in order to ensure a safe margin good magneto optic response for the experiment. The constructed Helmholtz coil was occasionally overdriven in current *(i.e. at maximum current of our power supply)*, generating a field strength at its center up to $B_c=30$ mT. The equation (1) gives,

$$B_c = \left(\frac{4}{5}\right)^{3/2} \frac{\mu_0 nI}{R} \qquad (1)$$

the strength of the magnetic field at the center of the Helmholtz coil $B_c$ with $n$ the number of turns per coil, $I$ the total current in series, $\mu_0$ is the permeability of free space and $R$ the inner radius of the Helmholtz coil. SI units are used in the equation. The ferrolens was positioned at the center of the field, center of the N-S axis of the Helmholtz coil *(see X-Axis in fig. 5a)*, **the experimental results are summarized in fig.6**.

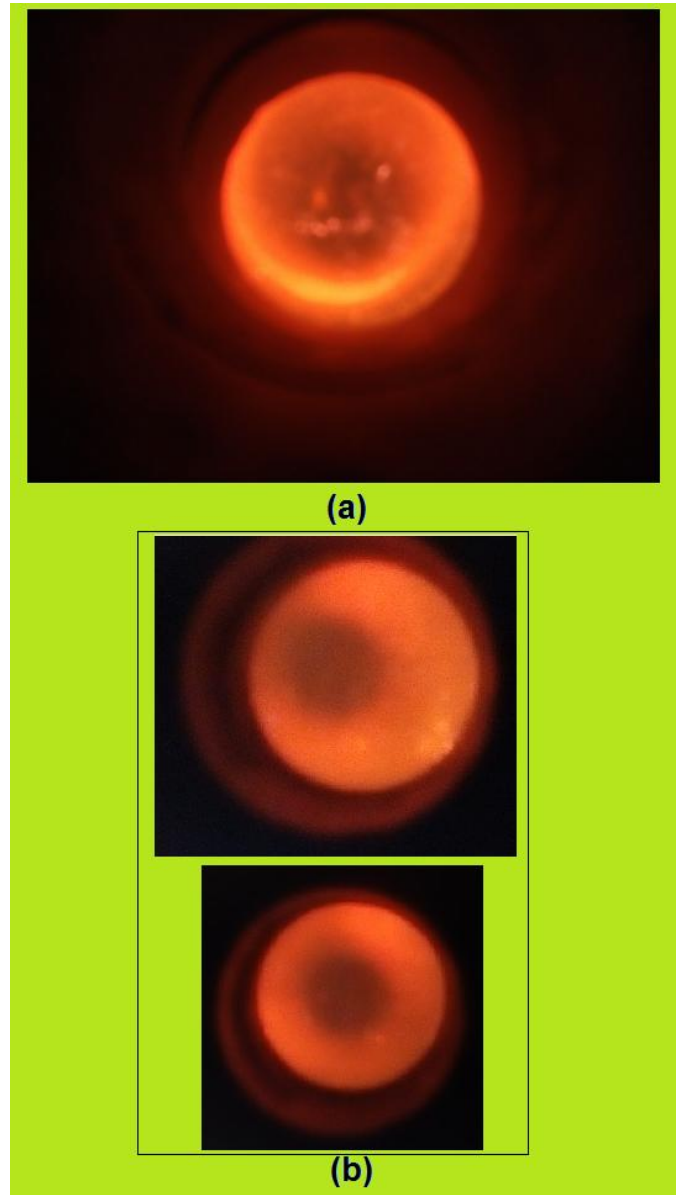

**Fig. 6 (a)** Homogeneous field of Helmholtz coil close-up at the center $B_c$ shown in real time by the ferrolens, field strength is 22mT. Some discrete flux lines are barely visible **(b)** Off-center observation of the field with the ferrolens.

At 20-30mT reaching its sensitivity threshold at 10mT, the ferrolens mini-Ferrocell, magnetic-optic response *falls exponentially*. Also, due the relative large size of the source (i.e. Helmholtz coil) compared to the ferrolens and the very straight *flux lines* of the field we could barely, very fade *record these* in fig.6a close-up photo and only at *a slight angle* are made visible. Else, the lines of force are piercing the lens perpendicular with minimum imprint and most of the time are completely black at the center and not discretely visible and only viewed as few bright dots whenever the camera is *directly facing* the lens. **We increased also the LED ring distance** from the lens from 2cm to 6cm to dim the light which helped to see discrete flux lines. The results, see fig.6a, comply with what is predicted by theory about the geometry of homogeneous magnetic fields generated inside Helmholtz coils [14] similar to the results we obtained in **Experiment-A** fig.2a of the N-S field of two separated attracting magnets and the field inside an air solenoid fig.2b [4]. The part of the total magnetic field of a Helmholtz which is also its homogenous field is located inside its inner diameter cylindrical volume. However even this, strictly speaking, is not ideal homogeneous which translates to same field strength $B$ everywhere thus same magnetic flux density but also zero curl inside the coil thus straight parallel flux lines, as shown by equation (2) of zero curl:

$$\nabla \times \mathbf{B} = 0 \qquad (2)$$

Although with the Hall sensor magnetometer we used in this experiment which has a resolution of one Gauss (i.e. 0.1 mT) we could not detect any variation in the magnetic flux density located at the total inner cross section area at the center of its X- Axis, see fig.5a, its curl varies slightly inside its total inner diameter cylindrical volume and is not zero everywhere as shown and predicted by theory in the Wolfram simulation of the curl of a Helmholtz coil field *shown in* fig.5b. Therefore, strictly speaking in ideal terms, only about a 25% theory predicts of its inner volume fulfils both conditions regarding an ideal homogenous magnetic field thus same flux density $B$ and almost zero curl magnetic flux (i.e. straight parallel lines of force). A special 3D printed artificial light funnel *(see fig.5a and also supplementary material [13])* which serves also as a dark box during day for increased contrast photography was constructed for illuminating the ferrolens and guiding it inside the Helmholtz coil.

**Experiment D – For the ferrolens positioned Off-center of the Helmholtz coil and its N-S axis perpendicular to the ferrolens glass surface (non-ideal homogeneous field):**

The idea of using magneto optic response of ferrofluids to find the symmetry center of non-ideal homogeneous magnetic fields used as a beam focus mechanism in acceleration particle physics, was first proposed in detail by J.K. Cobb and J.J. Muray in 1967 [15]. They used a cm thick vial filled with a ferrofluid solution. Certainly, this application and conclusions are applicable also for the much more sensitive μm thick thin film of the ferrolens. We demonstrated in **Experiment-D** how the ferrolens display corresponds to a non ideal homogeneous magnetic field region inside the Helmholtz coil by *displacing out of mechanical alignment* one of the two coils of the Helmholtz resulting the ferrolens to be OFF-center the dimensional mechanical X-Axis (see fig.5a) of the Helmholtz coil, outside its perfect 25% cylindrical volume (see fig.5b) ideally centered homogeneous field region. Where the field in this OFF- center region is not so much homogenous and a slightly curled and spatial varying density of magnetic flux exists, see fig.6b of the field display by the ultra sensitive ferrolens, the mini-Ferrocell. We displaced one of the coils in a diagonal direction from the center X-Axis shown in fig.5a and we see the resulted off-center field display in fig.6b compared to its symmetry center display.

The difference when compared with the on-center display of the ferrolens, see fig.6b (last photo), is immediately visible and obvious. In non ideal situations, the field symmetry center generated by a Helmholtz coil due to mainly construction imperfections does not always coincide perfectly with the mechanical X-Axis of the coil. Positioning the ferrolens inside the symmetry central area, most homogeneous field region *(see fig.5b)*, will generate a correspondingly more symmetrical and vivid display of field.

**Experiment E – For the ferrolens positioned at the center of a 63mm radius Helmholtz coil (homogeneous field):**

Experiments with Helmholtz coils and the ferrolens were carried out in both continents Europe (Greece) and America (USA) constructing two different Helmholtz coil setups and comparing the results. In this here **Experiment-E**, an almost double the radius and number of winding turns Helmholtz coil was designed and constructed by the authors. This allowed us to use a much larger ultra sensitive ferrolens of 62mm in diameter compared to the previously used ferrolens type, the 35mm mini-Ferrocell, fitted with a 42mm in perimeter LED artificial lighting ring positioned a couple of cm below of the lens and at its center *(see supplementary material and Appendix of this paper [13] for more technical details)*.

This particular experimental setup allowed us to see with the ferrolens a more *zoomed out* display of the field inside the Helmholtz coil and also how sensitive the ferrolens display is on magnetic field strength variations.
The results of this experiment are summarized in **fig.7**.
In fig.7(a) and fig.7(b) we see the obvious different magnetic optic response of the ferrolens with the lens being positioned parallel to the field lines inside the Helmholtz coil at its center. The flux lines in this configuration are skimming the surface of the lens. We observe as previously, that as we approach the sensitivity threshold of the ferrolens its magneto optic response and especially the *brightness* of the display, falls exponentially as shown by the comparison, since only a 8mT (i.e. 22mT to 30mT) difference in the field yielded a rather large visible response difference.

**Fig.7(c)&(d)** show the results of the experiment but this time the ferrolens is positioned perpendicular inside the Helmholtz to the magnetic moment, N-S axis, of the coil. These results are consistent with the previously experimental results of fig.6. Most of the very straight flux lines of the Helmholtz field which is considerable more homogenous than the direct field of Experiment-**A** with the permanent magnets, because this fact, are going almost perfectly perpendicular through, piercing the lens thickness and therefore scatter no light and are displayed as a dark circle area at the center of the display and not being possible to be shown discretely on this experimental setup and low field values. Especially, not possible to show discrete flux lines at the center of the display in this configuration, with the LED light ring so close to the surface of the lens, only 2cm below the lens.

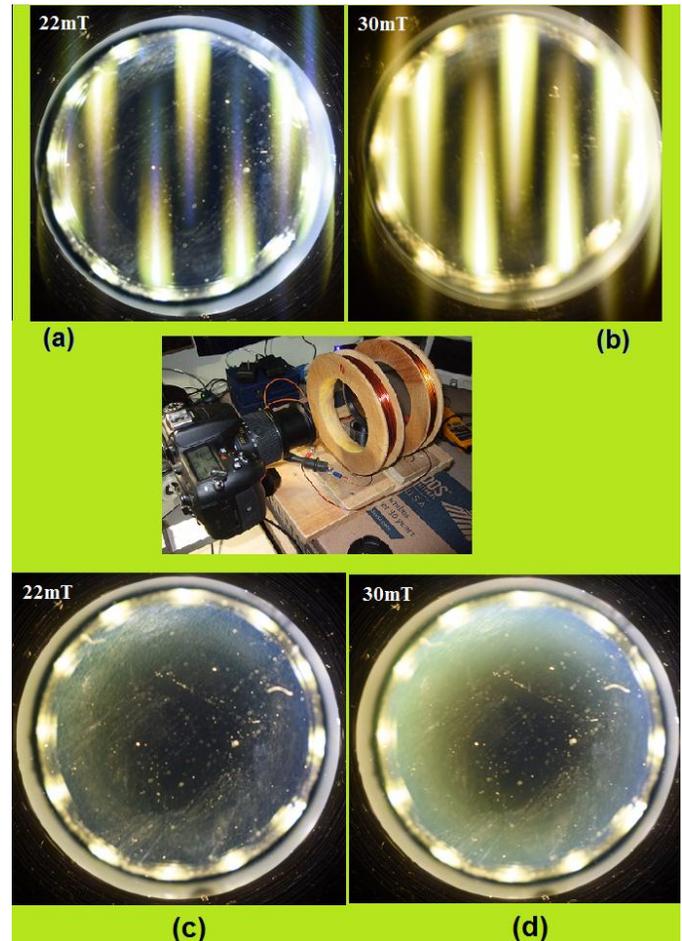

**Fig. 7** Ferrolens magnetic optical response on various Helmholtz coil magnetic field strengths and observation angles. **(a)** Bc=22mT field with the ferrolens glass surface parallel to the magnetic moment N-S axis of the field. The magnetic lines of force are skimming the surface of the lens. **(b)** The same as before but this time at Bc=30mT field strength. Improved magnetic optic response of the ferrolens. **(c)&(d)** Bc=22mT and 30mT field display response of the ferrolens but this time the lens is positioned with its glass surface perpendicular the magnetic moment, N-S axis of the magnetic homogeneous field. The results are consistent with the previous experiments of fig.6. In this particular view angle of the Helmholtz field looking in detail at the full size hi-resolution individual photographs which can be found in the supplementary material [13], we can distinguish discrete magnetic flux lines of the field on the perimeter of the ferrolens display imprinted as very thin and fade threats piercing trough the thickness of the lens.

Nevertheless, observing closely at the full size high resolution individual photographs of fig.7c&d found inside the supplementary material [13], we can see some discrete magnetic flux lines imprinted on the perimeter of the ferrolens display, similar to very thin threads piercing the ferrolens surface and aligned with magnetic moment of the Helmholtz coil. As before as in the case of fig6.a&b, we can see very evidently the magnetic optic response difference of the ferrolens display when going from 22mT to 30mT field strength. There also dust particles present in fig.7 photographs. An additional technique is proposed inside the Appendix [13] to make individual magnetic flux lines of the field more visible for demonstration purposes.

## 4. Conclusion

We have herein shown and proposed a potential new application made possible with the latest generation of ultra sensitive magneto optic response ferrolenses for displaying in real time on air homogeneous magnetic fields of 10mT strength and above. Sensitivity of the ferrolens is controlled by the thin film thickness and the degree of dilution of the nanoparticles inside the carbon-based colloidal fluid solution of the thin film.

We have probed various homogeneous magnetic field sources in our experiments such us the direct field between two unlike poles N-S of two separated permanent magnets, the field inside air solenoids and also a Helmholtz coils. Nevertheless, the potential applications of the ferrolens device in displaying homogeneous magnetic fields could include also other homogeneous field sources such us a MRI scanner [13] and is not specific only to these magnetic sources used in this research but can cover any type of macroscopic homogeneous magnetic field generated, as long as it has a minimum field strength of about 10 mT, necessary to trigger an adequate magneto optic response on the ferrolens.

The ferrolens when placed perpendicular to the straight parallel magnetic lines of force of a homogeneous magnetic field penetrating the thickness of the lens because of minimum light scattering on the aligned nanochains inside the thin film, the magnetic field lines are displayed as black discrete lines when white artificial lighting is used including also depth of field information and therefore producing an overall 3D tunnel effect display, fig.2a of the field.  The displayed field geometry agrees with that predicted by the known theory.

For comparison we examined also the display of the ferrolens of non-homogeneous fields such as that of single dipole permanent magnets, repulsing field between two like N-N poles. In the non-homogeneous fields cases, the force lines which have curl with an horizontal directional component to the ferrolens surface are displayed as illuminated bright colored lines fig.3&4 due to the increased light scattering in contrast to the vertical penetrating force lines near the poles that are displayed as black. In general vertical magnetic flux to the ferrolens surface is displayed as black whereas curled magnetic flux as bright illuminated colored lines.

We demonstrated also a possible mechanism and additional new information herein of how an incident non-homogeneous field is compressed inside the finite thickness of the thin film of the ferrolens retaining depth of field information and magnetic transparency and how different perspective views of the displayed field can be obtained by adjusting the incident magnetic field angle. fig.3&4.

The homogeneous magnetic field display of the ferrolens was further verified with two different configuration prototype Helmholtz coils, fig.5. and fig.7. Again observations agreed with the well known theoretical predictions of the homogeneous magnetic field geometry of Helmholtz coils. It was demonstrated how the ferrolens corresponds and can display a non-homogeneity inside a Helmholtz coil due to a mechanical misalignment of the two individual coils of the Helmholtz and therefore how it can be used to detect the symmetry magnetic homogeneous center region of a Helmholtz coil magnetic field. Therefore, a possible new application of the ferrolens in particle physics, beam magnetic focus technology and apparatus for finding any misalignment in the magnetic focus used for the beam. In conjunction with a spectrometer sensing the display output of the ferrolens, sub-millimetre alignment accuracy could be possible.

The operational limitations were described (see section 2) of this new sensitive ferrolens and also parameters affecting the longevity of the device. Especially based on our experiments, for the application of the latest generation of ultra sensitive ferrolenses in Helmholtz coils, we recommend a minimum of 20mT field strength for an adequate magnetic optic response.

**Notes:**

[1] Commercially known as a Ferrocell. Ferrocell®USA Trademark. US Patent 8246356 "Magnetic flux viewer".
Website: https://www.ferrocell.us/

[2] Ferrolens wiki:
https://en.everybodywiki.com/Ferrolens

[3] EFH1 ferrofluid product specifications:
https://tinyurl.com/e2nvs97u

Supplementary material [13],
https://tinyurl.com/dkjntac,  includes the Appendix of this paper.